\documentclass[a4paper,12pt]{article}
\usepackage[utf8]{inputenc}
\usepackage{amsmath}
\usepackage{epsf}
\usepackage{graphicx}
\usepackage{color}
\usepackage{indentfirst}
\usepackage{cite}

\begin{document}

\begin{center}
{\Large\textbf{On the search for the electric dipole moment of
strange and charm baryons at LHC and parity violating (P) and time
reversal (T) invariance violating spin rotation and dichroism in
crystal}}
\end{center}

\bigskip

\begin{center}
\textbf{V.G. Baryshevsky}
\end{center}

%

\begin{abstract}
In a bent crystal the $P$-odd effect of short-lived baryon spin
rotation could imitate spin rotation caused by assumed EDM.
Use of different behavior of $P$-odd  and $T$-odd spin rotations at crystal turning around the direction of particle momentum makes it possible to exclude P-odd rotation contribution, when measuring short-lived baryons EDM.
Subtraction  of the measurements  results for angle ranges $
\varphi $ and $ \varphi + \pi $ from each other enables measuring
$T$-odd rotation at scattering of negatively charged beauty  and
neutral baryons by axes of unbent crystal.
\end{abstract}

\noindent \textbf{\small{Crystal, bent crystal, charm baryon,
electric dipole moment, spin rotation, parity violation, magnetic
moment}}


\section{Introduction}

The spin rotation phenomenon of channeled particles, moving in a
bent crystal, which was theoretically predicted in \cite{b1} and
observed in \cite{b2,b3,b4}, gives us the opportunity to measure
anomalous magnetic moment of high energy short-lived particles.
The appearance of beams with energies up to $7~Tev$ on LHC and
further growth of particles' energy and beams' luminosity on FCC
do essentialy improve the possibility of using the phenomenon of
spin rotation of the high energy particles in bent crystals and
spin depolarization of such particles for measuring anomalous
magnetic moments of positively charged, as well as neutral and
negatively charged  short-lived hyperons, and $\tau$ -lepton
\cite{b5,b6,b7,b8}. The detailed analysis of conditions of the
experiment on measuring magnetic dipole moment (MDM) of charm
baryon $\Lambda^{+}_{c}$ on LHC, which has confirmed the
possibility of measuring MDM of such baryon on LHC, was accomplished

recently in \cite{b9}.
Strong electric field affects the channeled particle in a bent
crystal. As a consequence, the spin rotation phenomenon of the
channeled particle allows to obtain information about the possible
value of the electric dipole moment of short-lived baryons,
which elementary particles can obtain as a result of
the violation of the T-invariance \cite{b10}.

It should be noted, that besides electromagnetic interaction the channeled particle moving in a crystal experience weak interaction with electrons and nuclei as well as strong interaction with nuclei. Mentioned interactions lead to the fact, that in the analysis of the particle's spin rotation, caused by  electric dipole moment interaction with electric field, both $P (P,T)$ non-invariant spin rotation, resulting from weak interaction, and spin dichroism should be considered \cite{b11,b12}.

Let us concretize the mentioned above for the interested case of baryons moving in a bent crystal.

\section{Parity violating (P) and time reversal (T) invariance violating spin rotation and dichroism in crystal}

The spin rotation phenomenon for high-energy particles, moving in
a bent crystal, as a result of quasi-classical motion manner of
particles  channelled in crystals, can be described by equations
similar to those for motion of particles' spin in the storage ring
with the inner target \cite{b11,b12}.
The theory, which describes motion of the particle spin in
electromagnetic fields in a storage ring, has been developed in
many papers \cite{b13,b14,b15,b16,b17,b18,b19}.

According to \cite{b13,b14,b15,b16,b17,b18,b19},
the basic equation, which describes particle spin motion in an
electromagnetic field, is the Thomas-Bargmann--Michel--Telegdi
(T-BMT) equation. Refinement  of the T-BMT equation, allowing us
to consider the possible presence of the particle EDM, was made in
\cite{b20,b21}.

Now let us consider a particle with  spin $S$ which  moves in the
electromagnetic field.
The term ''particle spin'' here means the expected value of the
quantum mechanical spin operator $\hat{\vec{S}}$ (hereinafter the symbol marked with "hat"
means a quantum mechanical operator).

Spin motion is described by the Thomas--Bargmann--Michel--Telegdi
equation (T-BMT)as follows:
\begin{equation}
    \frac{d \vec{S}}{d t}=[\vec{S}\times\vec{\Omega}]\,,
    \label{eq1}
\end{equation}
\begin{equation}
    \vec{\Omega}=\frac{e}{mc}\left[\left(a+\frac{1}{\gamma}\right)\vec{B}
    -a\frac{\gamma}{\gamma+1}\left(\vec{\beta}\cdot\vec{B}\right)\vec{\beta}-
    \left(a+\frac{1}{\gamma+1}\right)\vec{\beta}\times\vec{E}\right]\,,
    \label{eq2}
\end{equation}
where $\vec{S}$ is the spin vector, $t$ is the time in the
laboratory frame,
$m$ is the mass of the particle, $e$ is its charge,  $\gamma$ is
the Lorentz-factor,
$\vec{\beta}=\vec{v}/c$, where $\vec{v}$ denotes the particle velocity, $\vec{E}$ and
$\vec{B}$ are the electric and magnetic fields at the point of
particle location in the laboratory frame, $a=(g-2)/2$ and $g$ is the gyromagnetic ratio
(by definition, the particle magnetic moment $\mu=(eg/2mc \hbar)S$, where $S$ is the particle spin).
%
The T-BMT equation describes the spin motion in the rest frame of
the particle, wherein the spin is described by the three component
vector $\vec{S}$.
In practice the T-BMT equation works well for the description of
spin precession in the external electric and magnetic fields
encountered in typical present--day accelerators.
Study of the T-BMT equation allows us to determine the major
peculiarities of spin motion in an external electromagnetic field.
However, it should be taken into account that particles in an
accelerator or bent crystal have an energy spread and move along different orbits.
This necessitates one to average the spin--dependent parameters of
the particle over the phase space of the particle beam.
This is why one must always bear in mind the distinction between
the beam polarization $ \vec{\xi} $ and the spin vector $\vec{S}$.
A complete description of particle spin motion can be made
applying spin density matrices equation (in more details see
\cite{b12,b22}).

If a particle possesses an intrinsic electric dipole moment, then
the additional term, describing spin rotation induced by the EDM,
should be added to (\ref{eq1}):
\begin{equation}
    \frac{ d \vec{S}_{\mathrm{EDM}}}{  d  t}=\frac{ d }{S\hbar}
    \left[\vec{S}\times\left\{(\vec{\beta}\times\vec{B}+\vec{E})-\frac{\gamma}{\gamma+1}\vec{\beta}(\vec{\beta}\vec{E})\right\}\right]\,,
    \label{eq3}
\end{equation}
where $d$ is the electric dipole moment of the particle.

As a result, the motion of particle spin due to the magnetic
and electric dipole moments can be described by the following
equation:
\begin{eqnarray}
    \frac{ d \vec{S}}{d t} & =& \frac{e}{mc}
    \left[ \vec{S}\times\left\{\left(a+\frac{1}{\gamma}\right)\vec{B}
    -a\frac{\gamma}{\gamma+1}\left(\vec{\beta}\cdot\vec{B}\right)\vec{\beta}
    - \left(a+\frac{1}{\gamma+1}\right)\vec{\beta}\times\vec{E}
    \right\} \right]\nonumber\\
    & &+ \frac{d}{\hbar
        S}\left[\vec{S}\times\left\{(\vec{\beta}\times\vec{B}+\vec{E})-
    \frac{\gamma}{\gamma+1}\vec{\beta}(\vec{\beta}\vec{E})\right\}\right]\,.
    \label{eq4}
\end{eqnarray}
%

\noindent Recall now, that electric and magnetic fields in a crystal are formed by atoms. Scattering on atoms leads to the fact, that the high-energy particle moving in a crystal experience interaction from electric and magnetic fields. However it is not only the electromagnetic interaction that influence on the scattering. Particles also participate in strong and weak interactions with electrons and nuclei.
It would be recalled that the particle refractive index in matter has the form:
\begin{equation}
    n=1+\frac{2\pi N }{k^{2}}f\left( 0\right)\,,
    \label{eq5}
\end{equation}
where $N $ is the number of scatterrers per cm$^{3}$ and $k$ is the
wave number of the particle incident on the target, $f(0)$ is the
coherent elastic zero--angle scattering amplitude.

Let us consider a relativistic particle refraction on the vacuum--medium boundary
(see \cite{b12} 
).

    %
The wave number of the particle in the vacuum is denoted $k$. The wave number of the particle in the  medium is $k^{\, \prime} = k n$. As is evident, the particle momentum in the vacuum $p=\hbar k~$ is not equal to the particle momentum in the medium.
Therefore, the particle energy in the vacuum $E=\sqrt{\hbar^2 k^2c^2+m^2 c^4}$ is not equal to the particle energy in the medium
$E_{\mathrm{med}}=\sqrt{\hbar^2 k^2 n^2 c^2+m^2 c^4}$.

The energy conservation law immediately requires the particle in medium to have the effective potential energy $V_{\mathrm{eff}}$. This energy can be easily found from the relation
\[ E={E_{\mathrm{med}}}+V_{\mathrm{eff}}\,,\]
    i.e.,
    \begin{equation}
        V_{\mathrm{eff}}=E-E_{\mathrm{med}}=- \frac{2 \pi {\hbar}^2}{m
            \gamma} {N} \, f(E,0) = (2 \pi)^3 N \, T(E)\,,
        \label{eq6}
    \end{equation}
        \[f(E,0)=- (2 \pi)^2 ~\frac{E}{c^2 \hbar^2} ~T(E)= - (2 \pi)^2
    ~\frac{m \gamma }{\hbar^2}~ T(E)\,.
    \]
\noindent where  $T(E)$ is the T-matrix.

Due to periodic arrangement of atoms in a crystal the effective potential energy is a periodic function of coordinates of a particle moving in crystal \cite{b12}.

    \begin{equation}
        U(\vec{r})=\sum_{\vec{\tau}} U(\vec{\tau}) e^{i \vec{\tau}
            \vec{r}}\,,
        \label{eq7}
    \end{equation}
    where $\vec{\tau}$ is the reciprocal lattice vector of the
    crystal,
    \begin{equation}
        U(\vec{\tau})=\frac{1}{V} \sum_{j} U_{j0}(\vec{\tau})
        e^{-W_j (\vec{\tau})} e^{i \vec{\tau} \vec{r}_j}\,,
        \label{eq8}
    \end{equation}
\noindent here $V$ is the volume of the crystal elementary cell, $\vec{r}_j$
is the coordinate of the atom (nucleus) of type $j$ in the crystal
elementary cell and the squared $e^{-W_j (\vec{\tau})}$ is equal
to the thermal-factor (i.e., the Debye-Waller factor), well-known for
X-ray scattering:
    \begin{equation}
        U_{j0}(\vec{\tau}) =-\frac{2 \pi \hbar^2}{m \gamma}
        F_j{(\vec{\tau})}\,,
        \label{eq9}
    \end{equation}
\noindent where $F_j{(\vec{\tau})}=F_j{(\vec{k}^\prime-\vec{k}=\vec{\tau})}$ is the amplitude of elastic coherent scattering of
the particle by the atom, $\vec{k}$ is the wave vector of the
incident wave and  $\vec{k}^\prime$ is the wave vector of the
scattered wave.

It will be recalled that in the case of a crystal, the imaginary part of
the amplitude $F_{j}(0)$ does not contain the contribution from the total cross
section of elastic coherent scattering. The imaginary part of $F_{j}(0)$ in a
crystal is only determined by the total cross
sections of inelastic processes. This occurs because in a crystal, unlike in
amorphous matter, the wave elastically scattered at a non zero angle, due to
rescattering by periodically located centers, is involved in formation of a coherent wave propagating through the crystal.

Elastic coherent scattering of a particle by an atom is caused by
Coulomb interaction of the particle with the atom electrons
and nucleus as well as with  weak and strong nuclear interaction with the electrons and nucleus.
Therefore, the scattering amplitude can be presented as a sum of
two amplitudes:
    \begin{equation}
        F_j{(\vec{\tau})}=F_j^{\mathrm{Coul}}{(\vec{\tau})}+F_j^{\mathrm{ws}}{(\vec{\tau})}\,,
        \label{eq10}
    \end{equation}
where $F_j^{\mathrm{Coul}}{(\vec{\tau})}$ is the amplitude of particle
scattering caused by Coulomb interaction with the atom
(it contains contributions from the Coulomb interaction of the
particle with the atom along with the spin-orbit interaction with
the Coulomb field of the atom);
$F_j^{\mathrm{ws}}{(\vec{\tau})}$ is the amplitude of elastic coherent
scattering of the particle caused by weak and strong interaction,
(this amplitude contains the terms independent of the incident
particle spin along with the terms depending on spin of both the
incident particle, electrons and nucleus, in particular, spin-orbit
interaction). Therefore, $U(\vec{r})$ and $U(\vec{\tau})$ can also
be expressed:
    \begin{eqnarray}
        \begin{array}{l}
            U{(\vec{r})}=U^{\mathrm{Coul}}{(\vec{r})}+U^{\mathrm{ws}}{(\vec{r})}\,,
            \\
            U{(\vec{\tau})}=U^{\mathrm{Coul}}{(\vec{\tau})}+U^{\mathrm{ws}}{(\vec{\tau})}\,.
            \label{eq11}
        \end{array}
    \end{eqnarray}

Suppose a high energy particle moves in a crystal at a small angle
to the crystallographic planes (axes) close to the Lindhard angle
${\vartheta_L \sim \sqrt{U/E}}$ (in a relativistic case
${\vartheta_L \sim \sqrt{2U/E}}$), where $E$ is the energy
of the particle, $U$ is the height of the potential barrier
created by the crystallographic plane (axis).
This motion is determined by the plane (axis) potential $\hat{U}
(\vec{\rho})$, which could be derived from ${U} (\vec{r})$ by
averaging over the distribution of atoms (nuclei) in the crystal
plane (axis).
As a consequence for the potential of periodically placed axes we can write:
 \begin{equation}
 \hat{U}(\vec{\rho})=\sum_{\vec{\tau_{\bot}}} U(\vec{\tau_{\bot}},\tau_{z}=0) e^{i \vec{\tau_{\bot}}\vec{\rho}} ,
 \label{eq11+1}
 \end{equation}
\noindent z-axis of the coordinate system is directed along the crystallographic axis (let us note that this expression can be obtained if all the terms with $ \tau_{z}\neq0 $ are removed from the sum (\ref{eq7})).
Replacing summation over $\vec{\tau_{\bot}}$ by integration over $\vec{\tau_{\bot}}$ we obtain an expression for the potential of separate crystallographic axis.

To obtain the potential of periodically placed planes we have:
\begin{equation}
\hat{U}(x)=\sum_{\tau_{x}} U(\tau_{x},\tau_{y}=0,\tau_{z}=0) e^{i \tau_{x}x} ,
    \label{eq11+2}
    \end{equation}

Y,Z-plane of the system of coordinates is parallel to the chosen crystallographic planes family. After replacing summation over $\vec{\tau_x}$ by integration over $\vec{\tau_x}$ we obtain an expression for the potential of separate crystallographic plane.


As a consequence, the potential $\hat{U} (\vec{\rho})$ for a
particle channeled in a plane (or axis) channel or moving over
the  barrier at a small angle, close to the Lindhard angle, can be
expressed as a sum \cite{b12}:
    \begin{equation}
        \hat{U}
        (\vec{\rho})=\hat{U}^{\mathrm{Coul}}{(\vec{\rho})}+\hat{U}^{\mathrm{sp.-orb.}}{(\vec{\rho})}+\hat{U}_{\mathrm{eff}}^{\mathrm{ws}}
        {(\vec{\rho})}+ \hat{U}^{\mathrm{mag}}{(\vec{\rho})}\,,
        \label{eq12}
    \end{equation}
where $\hat{U}^{\mathrm{Coul}}$ is the potential energy of particle Coulomb
interaction with the crystallographic plane (axis),
$\hat{U}^{\mathrm{sp.-orb.}}$ is the energy of spin-orbit interaction with
the Coulomb field of the plane (axis) and spin-orbit nuclear
interaction with the effective nuclear field of the plane (axis),
$\hat{U}_{\mathrm{eff}}^{\mathrm{ws}}$ is the effective potential energy of
weak and strong interaction of the incident particle with the
crystallographic plane (axis).
    %
    %
According to (\ref{eq9}) this part of the potential is determined by the amplitude of elastic
coherent forward weak and strong scattering $\hat{f}_{ws}(0)$, which depends on the incident
particle spin and electron and nucleus polarization,
$\hat{U}^{\mathrm{mag}}{(\vec{\rho})}$ is the energy of magnetic
interaction of the particle with electrons (nuclei).


Let us note that for high-energy particles in a quasi-classical approximation terms $\hat{U}^{Coul}$, $\hat{U}^{\mathrm{sp.-orb.}}$ and $\hat{U}^{mag}$ lead to BMT equations.

Thus when describing the particle motion in crystals, the contribution of weak and strong
interactions to formation of the effective potential acting on a particle from the
crystallographic planes (axes) should be taken into account along with the electromagnetic interaction.

In the case of high energies, when  the particle motion in the
potential $\hat{U}(\rho)$ can be described in the quasiclassical
approximation, the spin--evolution equations for a particle moving
in straight and bent crystals in the presence of the contribution
from $\hat{U}_{\mathrm{eff}}^{\mathrm{ws}}(\vec{\rho})$ appear to
be similar to those for  particle motion in a storage ring \cite{b12,b23}.

For the explicit expression for $U^{ws}_{eff}$ and the amplitude $\hat{f}^{ws}(0)$ in the presence of parity nonconservation and time (T) reversal violation see \cite{b11,b12}.

So all spin phenomena discussed for the case of storage ring with an inner target also occur in the case under
consideration \cite{b11,b12,b23}.
The energy of electrons-nuclei interactions in atoms can be neglected during the analysis of particle scattering on the atom at high energies. As a result, electrons' and nuclei contribution to the effective interaction energy can be considered separately. This allows us to express contribution of strong and weak interactions to the effective interaction energy as follows:
        \begin{equation}
            \label{eq14}
            \hat{U}^{str. w}_{eff}= -\frac{2\pi{\hbar^{2}}}{m\gamma} \left(  N_{e}( \vec{\rho}) \hat{f}_{ew}(0) + N_{nuc}( \rho) \hat{f}^{sw}_{nuc}(0)
            \right).
        \end{equation}
Here $ N_{e}(\vec{\rho})$ is the electron density in the point
$\vec{\rho}=(x,y)$ of the crystallographic plane (axis). Vector
$\vec{\rho}$ is orthogonal to the chosen plane (axis) family,
$N_{nuc}(\vec{\rho})$ is the nuclei density in the point $\rho$,
$\hat{f}_{ew}(0)$ is the amplitude of hyperon elastic coherent
scattering by an electron, caused by weak $P$ and $T$ violating
interactions, $\hat{f}^{sw}_{nuc}(0)$ is the amplitude of hyperon
elastic coherent scattering by a nucleui, caused by both strong
and weak, $P$ and $T$ violating interactions.

Parity nonconservation and time reveal violation lead to the dependence of $\hat{f}_{ew}$ and  $\hat{f}^{sw}_{nuc}$ on the spin orientation of colliding particles. As a result, $U^{sw}_{eff}$ also depends on the spin orientation of colliding high-energy particles.
This fact leads to the appearance of the quasi-optic effects of
spin rotation and spin dichroism for high energy particles
\cite{b12}. Stated contributions to the rotation should be added
to equations (\ref{eq4}), that describe spin rotation of a
particle moving in an electric field \cite{b12,b23}.

Lets now return to the discussion of the experiment for detection
of the electric dipole moment of short-lived baryons moving in
electrostatic field of a bent crystal. In this case electrons and
crystal nuclei are unpolarized. As a result, the contribution to
the scattering amplitude, which depends on the baryon spin
orientation, is caused only by parity nonconservation:
    \begin{equation}
        \label{eq15}
        \hat{f}_{w}(0)= B_{0w}+B_{w}\vec{\sigma}\vec{n},
    \end{equation}
where $\vec{\sigma}=(\sigma_{x}, \sigma_{y}, \sigma_{z})$ is the Pauli matrix describing baryon spin $\hat{\vec{S}}=\frac{1}{2}\hbar\vec{\sigma}$, $\vec{n}$ is the unit vector in particle momentum direction.
As a consequence, the contribution to effective interaction
energy, which is dependent on the baryon spin, can be written as:

    \begin{equation}
    \label{eq16}
    \hat{U}^{w}_{eff}= -\frac{2\pi\hbar^{2} }{m\gamma}\left(N_{e}(\rho)\hat{f}_{we}(0)+N_{nuc}(\vec{\rho})\hat{f}_{w nuc}(0)   \right),
    \end{equation}
where $N_{e (nuc)}(\vec\rho)$ is the density of electrons (nuclei)
in the point $\vec\rho$ of the plane (axis), $\hat{f}_{we}(0)$ is
the amplitude of baryon forward scattering by an electron, which is
cased by weak interaction, $\hat{f}_{w nuc}(0)$ is the amplitude
of baryon forward scattering by a nucleus, which is cased by
weak interaction. Therefore the spin-dependent part of effective
interaction energy $\hat{U}^{w}_{eff}$ reads according to
(\ref{eq15}) as follows :
    \begin{equation}
    \label{eq17}
    \hat{U}^{w}_{eff}= - \frac{2\pi\hbar^{2}}{m\gamma}\left( N_{e}(\rho)B_{e}+N_{nuc}(\rho)B_{nuc}\right) {\sigma}\vec{n} ,
    \end{equation}
where $B_{e}$ describes parity violating contribution caused by baryon interaction with electrons,
$B_{nuc}$ is for parity violating contribution caused by baryon interaction with nuclei.

As a result, due to parity violation  ${U}_{eff}$ contains the term, which depends on spin orientation  and is proportional to $\vec{\sigma}\vec{n}$. This term is similar to those  describing interaction of magnetic moment $\mu$ with magnetic field  $\mu\vec{\sigma}\vec{B}$ and electric dipole moment of baryon $ d $ with electric field $d\vec{\sigma}\vec{E}$.
When the baryon spin is directed along the  momentum, the effective energy reads:
    \begin{equation}
        \label{eq18}
        \hat{U}^{w}_{eff\uparrow\uparrow}= - \frac{2\pi\hbar^{2}}{m\gamma}\left( N_{e}(\vec{\rho})f_{we}^{\uparrow\uparrow}(0) + N_{nuc}(\vec{\rho})f_{w nuc}^{\uparrow\uparrow}(0)\right),
    \end{equation}
where $f_{we(nuc)}^{\uparrow\uparrow}(0) = B_{oe(nuc)}+B_{e(nuc)}$ is the amplitude of elastic coherent scattering of a particle with the spin parallel to its momentum.

When the baryon spin is directed oppositely to the  momentum, the effective energy reads:
    \begin{equation}
        \label{eq19}
        \hat{U}^{w}_{eff\downarrow\uparrow}= - \frac{2\pi\hbar^{2}}{m\gamma}\left( N_{e}(\rho)f_{we}^{\downarrow\uparrow}(0) + N_{nuc}(\vec{\rho})f_{w nuc}^{\downarrow\uparrow}(0)\right),
    \end{equation}
where $f_{we(nuc)}^{\downarrow\uparrow}(0) = B_{oe(nuc)} - B_{e(nuc)}$ is the amplitude of elastic coherent scattering of a particle with the spin antiparallel to its momentum.
\noindent Therefore, $\hat{U}^{w}_{eff\uparrow\uparrow}\neq\hat{U}^{w}_{eff\downarrow\uparrow}$.


Difference between these two energies, similar to the difference in energies of magnetic moment interaction with magnetic field  at spin parallel and antiparallel orientations to the field, determines the frequency $ \Omega_{w} $ of spin precession around particle momentum as follows:

    \begin{equation}
        \label{eq20}
        \Omega_{w}=\frac{Re U^{w}_{eff\uparrow\uparrow} -Re U^{w}_{eff\downarrow\uparrow} }{\hbar} ,
    \end{equation}
i. e.
    \begin{equation}
        \label{eq21}
        \Omega_{w}=- \frac{4\pi\hbar}{m\gamma}\left( N_{e}(\vec{\rho})Re B_{e}+N_{nuc}(\vec{\rho})Re B_{nuc}\right) ,
    \end{equation}

Note the amplitude $f(0)$ (amplitudes $B_{e}$ and $B_{nuc}$)
includes factor $\gamma$ (see (\ref{eq6})). Thus, the ratio
${\frac{1}{\gamma}} B_{e(nuc)}=B_{e(nuc)}^{'}$ does not include
this factor explicitly. Therefore the precession frequency reads

     \begin{equation}
                  \Omega_{w}=\frac{4\pi\hbar}{m}\left( N_{e}(\vec{\rho})Re B_{e}^{'}+N_{nuc}(\rho)Re B_{nuc}^{'}\right) ,
                  \label{eq22}
     \end{equation}
and this expression looks identical to nonrelativistic expression for frequency of spin precession.

%
For low energy neutrons the similar expression describes neutron
spin rotation caused by the weak parity violating interactions
\cite{b24,b25,b26}. This quasioptic effect is well studied for
neutrons.


\begin{figure}[htb]
  \centering
  \includegraphics[width=13 cm]{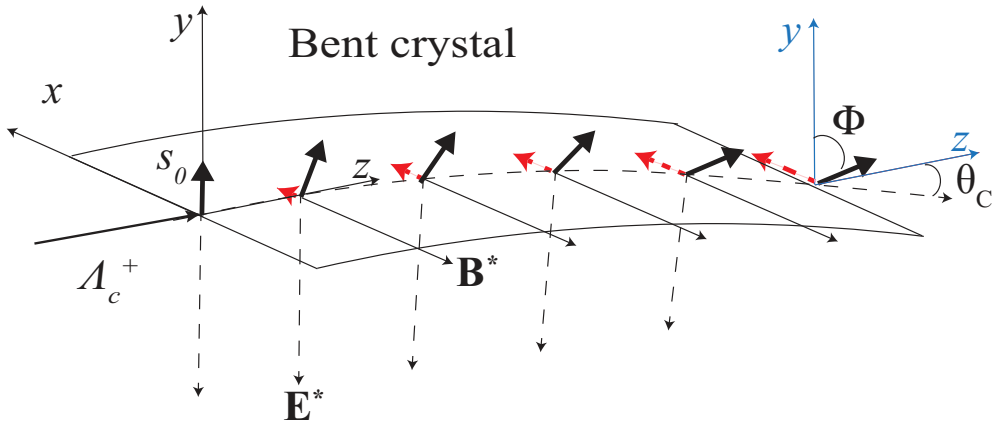}\\
  \caption{Behavior of the spin rotation caused by magnetic moment and EDM. The figure is reprinted from figure 2 (right)  in \cite{b10}. Black arrows represent spin rotation caused by magnetic dipole moment, red arrows represent spin rotation caused by electric dipole moment.}
    \label{fig1}
\end{figure}


\begin{figure}[htb]
  \centering
  \includegraphics[width=13 cm]{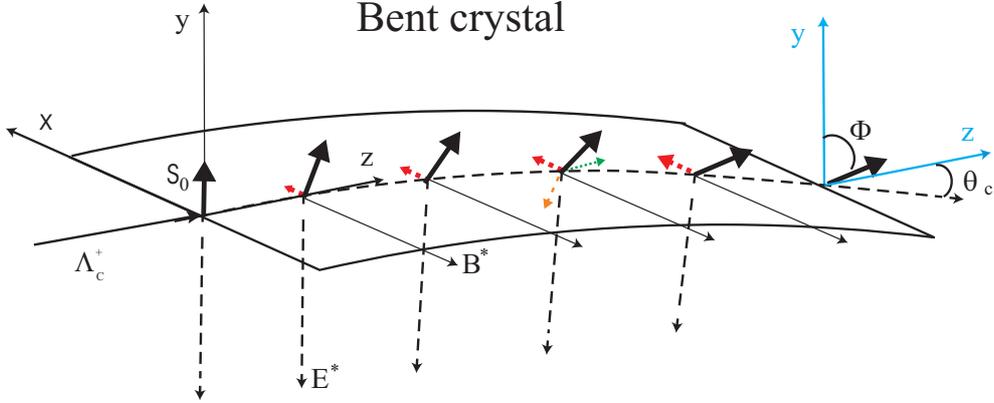}\\
  \caption{Behavior of the spin rotation caused by magnetic moment, EDM and P-violation spin rotation. Black arrows represent spin rotation caused by magnetic dipole moment, red arrows represent spin rotation caused by electric dipole moment, orange and green arrows represent additional rotation caused by P-violation.}
    \label{fig2}
\end{figure}



Note the amplitudes $B_{0}$ and $B_{nuc}$, which describe the amplitude of forward elastic coherent scattering, are complex.
Their imaginary parts are defined by the total interaction cross-section via optical theorem

    \begin{equation}
        \label{eq23}
    Im B_{0 e (nuc)} = \frac{k}{4\pi} \frac{\sigma^{e(nuc)}_{\uparrow\uparrow}+\sigma^{e(nuc)}_{\downarrow\uparrow}}{2};
    Im B_{e (nuc)} = \frac{k}{4\pi} \frac{\sigma^{e(nuc)}_{\uparrow\uparrow}-\sigma^{e(nuc)}_{\downarrow\uparrow}}{2},
    \end{equation}
where $\sigma^{e(nuc)}_{\uparrow\uparrow}$ is the total cross-section of weak interaction for the baryon, which spin is parallel to the electron (nucleus) momentum;    $\sigma^{e(nuc)}_{\downarrow\uparrow}$ is the total cross-section of weak interaction for the baryon, which spin is antiparallel to the electron (nucleus) momentum.

%
One should remember that for a crystal, where the atoms are
periodically located, the total cross-section of coherent elastic
scattering (i.e. scattering, which does not change the state of
scatterers) should be subtracted from the total cross-section of
particle scattering by an atom. The waves scattered elastically
interfere with the incident wave and cause no absorption.

%
%
Since the amplitude is complex, the operator $\hat{\Psi}(t)$ is non-Hermitian. Due to complexity of the effective energies $ {U}_{\uparrow\uparrow(\downarrow\uparrow)}$, baryon absorption in crystal depends on its spin orientation owing to baryon weak interaction with electrons and nuclei. The spin projection onto momentum direction changes, and even when the particle spin is orthogonal to the momentum direction, the spin component parallel to the momentum arises at particle propagation in the crystal (Figure \ref{fig2})!


The additional term caused by weak interaction in equations (\ref{eq4}), which describe spin rotation, can be obtained by the following approach \cite{b12}. The spin wave function $\hat{\Psi}(t)$ meets the equation as follows:
    \begin{equation}
            i \hbar \dfrac{\partial|\Psi(t)\rangle}{\partial t}= \hat{U}_{eff}|\Psi(t)\rangle.
            \label{eq24}
    \end{equation}

\noindent Baryon polarization vector $\vec{\xi}$ can be found via $|\Psi(t)\rangle$:
    \begin{equation}
               \vec{\xi}= \dfrac{\langle \Psi(t) |\vec{\sigma}|\Psi(t)\rangle}{\langle \Psi(t)|\Psi(t) \rangle}.
                \label{eq25}
    \end{equation}
\noindent From (\ref{eq25}) it follows:
        \begin{equation}
                       \dfrac{d\vec{\xi}}{dt}=\Omega_{w}[\vec{\xi} \times \vec{n}] -g (\vec{n}-\vec{\xi}(\vec{\xi}\vec{n})),
                        \label{eq26}
        \end{equation}
where $ \Omega_{w} $ is defined by (\ref{eq21}) and

    \begin{equation}
                g=\frac{1}{2} c [(\sigma_{\uparrow\uparrow}^{e}-\sigma_{\downarrow\uparrow}^{e})N_{e}(\vec{\rho})+(\sigma_{\uparrow\uparrow}^{nuc}-\sigma_{\downarrow\uparrow}^{nuc})N_{nuc}(\vec{\rho})],
                \label{eq27}
    \end{equation}

\noindent  Expression (\ref{eq26}) should be added to (\ref{eq4}). For relativistic baryons, when $ (\gamma>>1) $, and in case of magnetic field absence (nonmagnetic crystal) equations (\ref{eq4}) become simpler. Thus the equation for spin rotation of a particle, which moves in a bent crystal reads as follows:
        \begin{equation}
        \label{eq28}
    \dfrac{d\vec{\xi}}{dt}=-\frac{e (g-2)}{2mc} [\vec{\xi}\times[\vec{n}\vec{E}]] +\frac{d}{\hbar} [\vec{\xi}\times\vec{E}]+\Omega_{w}[\vec{\xi} \times \vec{n}] -g (\vec{n}-\vec{\xi}(\vec{\xi}\vec{n})).
        \end{equation}

\noindent Let us evaluate the effect. Precession frequency
$\Omega_{w}$ is determined by the real part of the amplitude of
baryon weak scattering by an electron (nucleus). This amplitude
can be evaluated in the energy range of about W and Z bosons
production and smaller by Fermi theory \cite{b26+}:
        \begin{equation}
            \label{eq29}
        ReB\sim G_{F}k=10^{-5}\frac{1}{m^{2}_{p}}k=10^{-5}\frac{\hbar }{m_{p} c}\frac{m}{m_{p}\gamma}=10^{-5}\lambda_{cp} \frac{m}{m_{p}\gamma} ,
        \end{equation}
where $ G_{F} $ is the Fermi constant, $m_{p}$ is the proton mass, $\lambda_{cp} $ is the proton Compton wavelength.
For particles with energy from hundreds of GeV to TeV $ReB\sim G_{F}k=10^{-16}$ cm.


For different particle trajectories in a bent crystal precession frequency $ \Omega_{w} $ could vary in the range $ \Omega_{w}\simeq 10^{3}\div 10^{4} sec^{-1}$. Therefore, when particle a passes 10~cm in a crystal, its spin undergoes additional rotation around momentum direction at angle $ \vartheta_{p} \simeq 10^{-6}\div 10^{-7}$ rad.


Absorption caused by parity violation weak interaction also contributes to change of spin direction. This rotation is caused by imaginary part of weak scattering amplitude and is proportional to the difference of total scattering cross-sections  $ \sigma_{\uparrow\uparrow} $ and $ \sigma_{\downarrow\uparrow} $ (see in (\ref{eq26},\ref{eq28}) the terms proportional to $g$).

    %

This difference in its turn is proportional to the factor, which is determined by interference of coulomb and weak interactions for baryon scattering by an electron, and  of strong (coulomb) and weak interactions for baryon scattering by nuclei.
\begin{equation}
 \sigma_{\uparrow\uparrow (\downarrow\uparrow)}=\int| f_{c(nuc)}+B_{0w}\pm B_{w}|^{2} d\Omega ,
  \label{eq30}
\end{equation}

\begin{equation}
        \sigma_{\uparrow\uparrow} - \sigma_{\downarrow\uparrow}= 2\int[(f_{c(nuc)}+B_{0w})B^{*}+( f_{c(nuc)}+B_{0w})^{*}B]d\Omega .
\label{eq31}
\end{equation}


\noindent When baryon trajectory passes in the area, where collisions with
nuclei are important (this occurs in the vicinity
of potential barrier for positively charged particles),  the value
$ g \sim 10^{6}\div10^{7} s^{-1}$. Multiple scattering also
contributes to spin rotation \cite{b12}. Particularly,  due to
interference of weak and coulomb interactions the root-mean-square
scattering angle appears changed and dependent on spin orientation
with respect to the particle momentum direction.


When measuring T-odd spin rotation in the electric field of a bent crystal, one can eliminate parity violating rotation by the following way. T-odd and P-odd spin rotations differently depend on crystal turning at $ 180^{\circ} $ around the direction of incident baryon momentum. Namely, P-odd effect does not change, while the sign of T-odd spin rotation is flipped due to change of the electric field direction. Subtracting results of measurements for two opposite crystal positions on could obtain the angle of rotation, which depends on T-odd effect only.  

\section{P and T-odd spin rotation in unbent crystals}


The angle of spin rotation $ \vartheta_{p} $ increases
significantly for a baryon in an unbent crystal, when particles
move at a small angle with respect to a crystal axes, since in
this case the scatterers density grows. Therefore, even for
short-lived  beauty (bottom) baryons with negative charge, the
weak amplitude $ReB$ can be measured. EDM interaction with
electric field of crystal axis has influence on the spin rotation
too.

    %
Study of EDM contribution to spin rotation of a particle, which
moves at a small angle with respect to a crystal axes, is hampered
by depolarization effect \cite{b7, b8, b12}.
%
%
Trajectories of the scattered particles, which azimuth angles are
in the vicinity  $ \varphi $ and $ \varphi + \pi $ ($ z $ axes is
directed along the crystal axes) contributes to EDM-caused spin
rotation with opposite signs due to different electric field
signs. At the same time, the $P$-odd rotation occurs around the
momentum direction and is not affected by the electric field
direction. Thus, $P$-odd rotation can be observed
in both bent and unbent crystals.
%
The T-odd spin rotation can be observed in unbent crystal if we use subtraction  of the measurements  results for angle ranges $
\varphi $ and $ \varphi + \pi $ from each other.
Such procedure leads to summation of contributions from $T$-odd
rotation.
Simultaneous measurement of spin orientation for all $ \varphi $
values (as well as for all polar angles for scattering by the
axes) provides intensity increase. For unbent crystals the same
measuring procedure  enables to use crystal with higher nucleus
charge that also contributes to effect increase. All the above is
primarily significant for negatively charged beauty baryons.

%
The similar reasoning is valid for measuring the anomalous
magnetic moment by means of axial scattering in unbent crystals.
Subtraction procedure can be applied for measuring both anomalous
magnetic moment and EDM  of neutral charm and beauty short-lived
baryons.

All the above immediately follows from the general expression for amplitude of elastic coherent scattering of a spin $1/2$ particle by a spinless (unpolarized) nuclei in presence of electromagnetic, strong and $P$-, $T$-odd weak interactions.
    \begin{equation}
   \hat{f}(\vec{q})=A(\vec{q})+B(\vec{q}) \vec{\sigma}\vec{N}+B_{0w}(\vec{q})+B_{w}(\vec{q})\vec{\sigma}\vec{N_{w}}+B_{T}\vec{\sigma}\vec{N_{T}},
     \label{eq32}
    \end{equation}
where $ A(\vec{q}) $ is spin-independent part of scattering amplitude, which is caused by electromagnetic and strong interactions.
$ \hbar\vec{q}=\hbar\vec{k}^{'} - \hbar\vec{k} $ is the transmitted momentum, $\hbar\vec{k}^{'}$ is the momentum of the scattered particle, $  \hbar\vec{k} $ is the momentum of the incident baryon,
 $\vec{k}^{'},\vec{k}  $ are the wave vectors,
$\vec{N}=\frac{[\vec{k}^{'}\times\vec{k}]}{[\vec{k}^{'}\times\vec{k}]}  $,
$\vec{N_{w}}=\frac{\vec{k}^{'}+\vec{k}}{|\vec{k}^{'}+\vec{k}|} $,
$\vec{N_{T}}=\frac{\vec{k}^{'}-\vec{k}}{|\vec{k}^{'}-\vec{k}|} $.

 The term, which is proportional to $ \vec{\sigma}\vec{N} $, is responsible for spin-orbit interaction contribution to scattering process. For electromagnetic interaction this contribution is determined by the particle magnetic moment.

T-odd part of scattering amplitude, which is proportional to  $ \vec{\sigma}\vec{N_{T}} $, in case of electromagnetic interaction is determined by electric dipole moment. T-odd nuclear interactions also contribute, when particles are scattered by nuclei.

Note, the expression (\ref{eq32}) is valid for both nonrelativistic and relativistic cases \cite{b12}.

With amplitude $ \hat{f}(\vec{q}) $ one can find the  cross-section of particle scattering by a crystal and polarization vector of the scattered particle.
According to \cite{b8} the scattering cross-section for a thin crystal reads:

\begin{equation}
    \label{eq33}
    \frac{d\sigma_{cr}}{d\Omega}=\frac{d\sigma}{d\Omega}\left\{(1-e^{-\overline{u^2}
        {q^2}}) +\frac{1}{N}\left|\sum_n e^{i\vec q\vec{r}_n^0}\right|^2 e^{-\overline{u^2} {q^2}}\right\},
\end{equation}
where $\vec{r}_n^0 $ is the coordinate of the center of gravity of the crystal  nucleus, $\overline{u^2}$ is the mean square  of thermal oscillations of nuclei in the crystal. The first term  describes incoherent scattering and the second one describes the coherent due to periodic arrangement of crystal nuclei (atoms). This contribution leads to the increase in the cross section.  This expression can be used as long as the  crystal length satisfies the  inequality $k(n-1) L\ll 1$, where $n$ is the particle refractive index in the crystal \cite{b8,b12}.
For thick crystals the influence of  channeling and depolarization effects become important.

\begin{equation}
    \label{eq34}
    \frac{d\sigma}{d\Omega}= tr\rho\hat{f^{+}}(\vec{q})\hat{f}(\vec{q}),
\end{equation}

\noindent where $ \rho $ is the spin density matrix of the incident particle.

 The polarization vector of a particle that has undergone a single scattering event can be found using the following expression: 
 \begin{equation}
    \label{eq35}
    \vec\xi= \frac{\mbox{tr}\rho f^+\vec\sigma f}{\mbox{tr}\rho f^+ f} =
    \frac{\mbox{tr}\rho f^+ \sigma f}{\frac{d\sigma}{d\Omega}}.
 \end{equation}

\noindent Using (\ref{eq32}) and  (\ref{eq35}) one can obtain the following expressions for polarization vector of the scattered particle and differential cross-section:
\begin{equation}
    \label{eq36}
\vec{\xi}=\vec{\xi_{so}}+\vec{\xi_{w}}+\vec{\xi_{T}},
\end{equation}
 where $ \vec{\xi_{so}} $ is the change of polarization vector due to spin-orbit interaction,
 $ \vec{\xi_{w}} $ is the change of polarization vector caused by weak parity violating   interaction, $ \vec{\xi_{T}} $ is the change of polarization vector caused by $T$-odd interaction.

\begin{eqnarray}
\label{eq37}
 \vec{\xi_{so}} & = & \left\{(|\overline{A}|^2 - |B|^2) \vec\xi_0 + 2
|B|^2 \vec N (\vec N\vec\xi_0)+ \right. \nonumber \\
&  & \left.  +  2 \texttt{Im} (\overline{A}B^*)[\vec N\vec\xi_0] +2 \vec N \texttt{Re} (\overline{A}B^*)\right\}\cdot \left(\frac{d\sigma}{d\Omega}\right)^{-1}, \nonumber \\
\vec{\xi_{w}}  & = &  \left\{(|\overline{A}|^2 - |B_{w}|^2)
\vec\xi_0 + 2
|B_{w}|^2 \vec{N_{w}} (\vec{N_{w}}\vec\xi_0)+ \right. \nonumber \\
 &  & \left.  + 2 \texttt{Im} (\overline{A}B^*_{w})[\vec{N_{w}}
\vec\xi_0]+2 \vec{N_{w}} \texttt{Re}
(\overline{A}B^*_{w})\right\}\cdot
\left(\frac{d\sigma}{d\Omega}\right)^{-1}, \nonumber \\
\vec{\xi_{T}} & = & \left\{(|\overline{A}|^2 - |B_{T}|^2)
\vec\xi_0 + 2
|B_{T}|^2 \vec{N_{T}}  (\vec{N_{T}} \vec\xi_0)+ \right. \nonumber \\
&  & \left.  +  2 \texttt{Im} (\overline{A}B^*_{T})[\vec{N_{T}}
\vec\xi_0]+2 \vec{N_{T}}  \texttt{Re}
(\overline{A}B^*_{T})\right\}\cdot
\left(\frac{d\sigma}{d\Omega}\right)^{-1},
\end{eqnarray}

where $\overline{A}= A+B_{0w} $. \noindent The differential
cross-section reads as follows:
\begin{eqnarray}
    \label{eq38}
& & \frac{d\sigma}{d\Omega}=\mbox{tr}\rho f^+ f = \nonumber \\
& &   |\overline{A}|^2 + |B|^2+|B_{w}|^2+|B_{T}|^2 +
2Re(\overline{A}B^*)\vec N \vec\xi_0 + \nonumber \\
& & 2Re(\overline{A}B_{w}^*)\vec{N_{w}} \vec\xi_0+
2Re(\overline{A}B_{T}^*)\vec {N_{T}} \vec\xi_0.
\end{eqnarray}

While derivating expressions (\ref{eq37}) and (\ref{eq38}) small terms containing production of $ B_{w} $ and $ B_{T} $, which are much smaller comparing to other terms,  were omitted.
%
\noindent As anticipated according to (\ref{eq37}) the angle
of polarization vector rotation for a baryon scattered in a
crystal is determined by rotations around three mutually
orthogonal directions.
%
%
%
Subtraction procedures for measurements, for which sign of vector products $ [\vec N \vec\xi_0] $ and $ [\vec {N_{T}} \vec\xi_0] $ change, enable statistics increase.
%
%
%
It is also useful to note that initially unpolarized particle beam
($\xi_0=0 $) in a crystal acquires polarization
directed along one of three vectors $\vec N$, $\vec{N_{w}}$, $\vec
{N_{T}}$, which carries information about all types of
interaction, namely: electromagnetic, strong and weak $P$-,
$T$-odd interactions.
%
%
According to (\ref{eq38}) amplitudes interference results in asymmetry in scattering caused by orientation of vectors $\vec {N_{T}}$, $\vec N$, $\vec{N_{w}}$ with respect to $\vec\xi_0, \vec{k}'$ and $ \vec{k}$ . Therefore, the scattered particles intensity is anisotropic.
%
%
Thus, measurement of angular distribution of intensity for a
particle beam scattered by crystal axes enables to obtain $T$-odd
contributions to the scattering cross-sections for short-lived
baryons. The same makes also possible to obtain $P$-odd and
spin-orbit contributions. Particularly, this can
be realized by the use of set of crystals with axes directed at
small angle with respect to momentum of the
scattered particles and detection of either nuclear reaction or
ionizing losses inside crystal detector.

\section{Conclusion}
%
In a bent crystal the $P$-odd effect of short-lived baryon spin
rotation could imitate spin rotation caused by assumed EDM.
%
%
Use of different behavior of $P$-odd  and $T$-odd spin rotations at crystal turning around the direction of particle momentum makes it possible to exclude P-odd rotation contribution, when measuring short-lived baryons EDM.
%
Subtraction  of the measurements  results for angle ranges $
\varphi $ and $ \varphi + \pi $ from each other enables measuring
$T$-odd rotation at scattering of negatively charged beauty  and
neutral baryons by axes of unbent crystal.
 %
 %
 The similar procedure can be applied for measuring both anomalous magnetic moment of the same particles.
 %
%
Measurement of $P$-odd rotation of short-lived baryons spin by crystal axes gives information about weak amplitude $B$.


\begin{thebibliography}{99}

\bibitem{b1} Baryshevsky  V.~G.,  Spin rotation of
ultrarelativistic particles passing through a crystal,
\emph{Pis'ma. Zh. Tekh. Fiz.} \textbf{5}, 3 (1979)  182--184.
\bibitem{b2} Baublis V.~V., et~al., First observation of spin precession of polarized $\Sigma^{+}$ hyperons channeled in bent crystals, LNPI Research Report, (1990--1991) E761 Collaboration (St. Pertersburg) (1992) 24--26.
\bibitem{b3} Chen D., Albuquerque I.~F. and Baublis V.~V.,  \emph{et~al.},
First observation of magnetic moment precession of channeled
particles in bent crystals,  \emph{Phys. Rev. Lett.} \textbf{69},
23  (1992) 3286--3289.
\bibitem{b4} Khanzadeev A.~V., Samsonov V.~M. Carrigan R.~A.  and  Chen D.,  Experiment to observe the spin precession of
channeled relativistic $\Sigma^{+}$ hyperons \emph{Nucl. Instr.
Methods B} \textbf{119}, 1-2 (1996) 266--270.
\bibitem{b5} Baryshevsky V.~G., Spin rotation and depolarization of high-energy particles in crystals at Hadron Collider (LHC) and Future Circular Collider (FCC) energies and the possibility to measure the anomalous magnetic moments of short-lived particles, (2015), arXiv:1504.06702.
\bibitem{b6} Baryshevsky V.~G.,\emph{Phys. Lett. B},\textbf{757}, (2016)  426--429.
\bibitem{b7} Baryshevsky V.~G., Depolarization of high-energy neutral particles in crystals and the possibility to measure anomalous magnetic moments of short-lived hyperons, (2016), arXiv:1608.06815.
\bibitem{b8} Baryshevsky V.~G., Spin rotation and depolarization of high-energy particles in crystals at LHC and FCC energies. The possibility to measure the anomalous magnetic moments of short-lived particles and quadrupole moment of $ \Omega $-hyperon, \emph{Nucl. Instr. Methods B},\textbf{402} (2017), 5--10.
\bibitem{b9} Bezshyyko O.~A., Burmistrov L., Fomin A.~S., et all, Feasibility of measuring the magnetic dipole moments of the charm baryons at the LHC using bent crystals, (2017), arXiv:1705.03382.
\bibitem{b10} Botella F.~J., Garcia Martin L.~M., Marangotto D., et all, On the search for the electric dipole moment of strange and charm baryons at LHC, (2016),  arXiv:1612.06769;  \emph{Eur. Phys J.C.} \textbf{77}, 181 (2017),  DOI 10.1140/epjc/s10052-017-4679-y.
\bibitem{b11}  Baryshevsky V.~G., Rotation of particle spin in a storage ring with a polarized beam and measurement of the particle
EDM, tensor polarizability and elastic zero-angle scattering amplitude, \emph{J. Phys. G}, 35 (2008).
\bibitem{b12} Baryshevsky V.~G.,High-Energy Nuclear Optics of Polarized Particles, World Scientific Publishing Company, 640 p. (2012).
\bibitem{b13} Derbenev Ja.~S. and Kondratenko, A.~M. (1973). Polarization kinetics of particles in storage rings, \emph{Zh. Eksp. Teor. Fiz.} \textbf{64}, 6, pp. 1918--1929
(\emph{Sov. Phys. JETP} \textbf{37} p. 968).
\bibitem{b14} Jackson, J.~D. (1976). On understanding spin--flip synchrotron
radiation and the transverse polarization of electrons in storage
rings,  \emph{Rev. Mod. Phys.} \textbf{48}, 3, pp. 417--433.
\bibitem{b15} Heinemann, K. and Hoffstaetter, G.~H. (1996).
Tracking algorithm for the stable spin polarization field in
storage rings using stroboscopic averaging, \emph{Phys.Rev.E}
\textbf{54}, 4, pp. 4240--4255.
\bibitem{b16} Mane, S.~R., Shatunov, Yu.~M. and  Yokoya, K. (2005). Spin--
polarized charged particle beams in high--energy accelerators,
\emph{Rep. Prog. Phys.} \textbf{68}, 9,  pp. 1997--2266.
\bibitem{b17} Hoffstaetter, G.~H., (2000). Polarized Protons in HERA,
LANL e-print arXiv:physics/0006007v1 [physics.acc-ph];
\emph{Nucl.Phys. A} \textbf{666} pp. 203--213.
\bibitem{b18} Hoffstaetter, G.~H. (2006). \emph{High Energy Polarized Proton Beams}
(A Modern View Series: Springer Tracts in Modern Physics, Vol.
\textbf{218}).
\bibitem{b19} Heinemann, K. and Barber, D.~P. (2001).
Spin transport, spin diffusion and Bloch equations in electron
storage rings, \emph{Nucl. Instrum. Methods A } \textbf{463},
1--2, pp. 62--67; Erratum \emph{Nucl. Instrum. Methods A }
\textbf{469}, 2,  p. 294 (2001).
\bibitem{b20} Ford, G.~W. and  Hirt, C.~W. (1961). University of Michigan Report unpublished.
\bibitem{b21} Mane, S.~R., Shatunov, Yu.~M. and  Yokoya, K. (2005). Spin--
polarized charged particle beams in high--energy accelerators,
\emph{Rep. Prog. Phys.} \textbf{68}, 9,  pp. 1997--2266.
\bibitem{b22}  Baryshevsky V.~G., Bartkevich A.~R., Tensor polarization of deuterons passing through matter,\emph{J. Phys. G}, \textbf{39}, (2012).
\bibitem{b23} Baryshevsky V.~G., Gurinovich A.~A., Spin rotation and oscillations of high energy particles in a crystal and possibility to measure the quadrupole moments and tensor polarizabilities of elementary particles and nuclei, \emph{Nucl. Instr. Methods B},\textbf{252}, (2006), 92.
\bibitem{b24} Michel F.~C.,Parity nonconservation in nuclei, \emph{Phys. Rev.},\textbf{133}, 2B,(1964), B329--B349.
\bibitem{b25} Forte M., Parity violation effects in neutron scattering and capture,\emph{Lett.Nuovo Cimento},\textbf{28}, 16, (1980), 538--540.
\bibitem{b26} Forte M., Heckel B.~R., Ramsey N.~F., Green K., Greene G.~L., Byrne J., Pendlebury J.~M., First measurement of parity–nonconserving neutron–spin rotation: The tin isotopes, \emph{Phys. Rev. Lett.},\textbf{45}, 26, (1980), 2088--2092.
\bibitem{b26+} Leader E., Predazzi E., An introduction to Gauge Theories and the "New Physics", Cambridge University press, (1982).
\end{thebibliography}
\end{document}